\documentclass[superscriptaddress,twocolumn,showpacs,showkeywords,preprintnumbers,amsmath,amssymb,aps,prb]{revtex4}
\usepackage{graphicx}% Include figure files
\usepackage{dcolumn}% Align table columns on decimal point
\usepackage{bm}% bold math
\begin{document}

\title{Developing and Researching PhET simulations for Teaching Quantum Mechanics}


\author{S. B. McKagan}
\affiliation{University of Colorado, Boulder, CO 80309, USA}

\author{K. K. Perkins}
\affiliation{University of Colorado, Boulder, CO 80309, USA}

\author{M. Dubson}
\affiliation{University of Colorado, Boulder, CO 80309, USA}

\author{C. Malley}
\affiliation{PixelZoom, Inc., Boulder, CO 80303, USA}

\author{S. Reid}
\affiliation{University of Colorado, Boulder, CO 80309, USA}

\author{R. LeMaster}
\affiliation{University of Colorado, Boulder, CO 80309, USA}

\author{C. E. Wieman}
\affiliation{University of British Columbia, Vancouver, BC V6T 1Z1, CANADA}
\affiliation{University of Colorado, Boulder, CO 80309, USA}

\date{January 17, 2008}

\begin{abstract}
Quantum mechanics is difficult to learn because it is counterintuitive, hard to visualize, mathematically challenging, and abstract.  The Physics Education Technology (PhET) Project, known for its interactive computer simulations for teaching and learning physics, now includes 18 simulations on quantum mechanics designed to improve learning of this difficult subject.  Our simulations include several key features to help students build mental models and intuitions about quantum mechanics: visual representations of abstract concepts and microscopic processes that cannot be directly observed, interactive environments that directly couple students' actions to animations, connections to everyday life, and efficient calculations so students can focus on the concepts rather than the math.  Like all PhET simulations, these are developed using the results of education research and feedback from educators, and are tested in student interviews and classroom studies.  This article provides an overview of the PhET quantum simulations and their development.  We also describe research demonstrating their effectiveness and share some insights about student thinking that we have gained from our research on quantum simulations.
\end{abstract}

\maketitle

\section{Introduction}

Quantum mechanics has challenged many of the greatest minds in physics, so it is no surprise that it is a difficult subject for students to learn.  In addition to the standard rigors associated with any topic in physics, quantum mechanics presents many of its own unique challenges that conspire to make it extraordinarily difficult and frustrating for most students to build mental models.  It is counterintuitive and surprising to find that the microscopic world does not behave at all the way we would expect, as the intuitions we have built up from interacting with our daily environment do not hold up.  Because most of the phenomena we study in quantum mechanics cannot be observed directly, it is often difficult to construct mental models by which to visualize such elusive phenomena.  It is also mathematically challenging, involving lengthy calculations to analyze the simplest phenomena, with most real-world phenomena falling outside the realm of our ability to calculate.  Finally, at least in the form it is often taught, quantum mechanics is disconnected from everyday life, focusing on simplified abstract models at worst, and phenomena with which we have no direct experience at best.

Extensive research in quantum mechanics education shows that students often do not learn what instructors would like them to learn in high school modern physics courses~\cite{Petri1998a,Mashhadi1999a,Harrison2000a,Olsen2002a,Taber2005a}, sophomore level modern physics courses~\cite{Steinberg1996a,Wittmann2002a,Wittmann2005a,McKagan2006a,Vokos2000a,Bao2002a}, junior level quantum mechanics courses~\cite{Vokos2000a,Bao2002a,Johnston1998a,Singh2001a,Sadaghiani2006a,Brookes2006a}, and even graduate courses~\cite{Singh2006a}.  Research on the development of transformed modern physics courses~\cite{Muller2002a,Zollman2002a,McKagan2007a}, as well as on tutorials targeting specific student difficulties~\cite{Ambrose2005a,Wittmann2005b,Wittmann2006a,Singh2006b}, suggests that improved student learning is possible.  However, most research that has been done so far has focused on only a few key topics such as atomic models, the photoelectric effect, and the properties of wave functions.  This is only the tip of the iceberg, and much remains to be discovered regarding student learning of other topics, including non-traditional topics such as real-world applications and interpretations of quantum mechanics, as well as student beliefs about quantum mechanics and the nature of science, and how to best improve student understanding of all aspects of quantum mechanics.  There is a need for further research and development of techniques and tools for effectively teaching quantum mechanics.

Educational computer simulations are promising tools that have been shown to be effective in helping students learn many topics in introductory physics~\cite{Finkelstein2005a,Adams2008a,Adams2008b}.  Because of the added problems of visualizing and building an intuition for the abstract principles of quantum mechanics, the power of simulations to provide interaction, visualization, and context has the potential to be even more helpful in this subject than in introductory physics.

Many teachers and researchers have developed computer simulations to assist students in learning quantum mechanics.~\cite{Goldberg1967a,Brandt1995a,Hiller1995a,Thaller2000a,Belloni2006a,vqm,qpo,falstad}  While many of these simulations may be useful for providing visual models of quantum phenomena, research on their user interface and effectiveness for learning has been limited.  Many of the user interfaces or representations of physics are not consistent with research on user-interface design~\cite{Adams2008b} and how students learn~\cite{Bransford1999a}, potentially limiting their effectiveness.

The Physics Education Technology (PhET) Project creates research-based interactive computer simulations for teaching and learning physics and makes them freely available from the PhET website (http://phet.colorado.edu).  The simulations are animated, interactive, and game-like environments where students learn through exploration.  We emphasize the connections between real-life phenomena and the underlying science, and seek to make the visual and conceptual models of expert physicists accessible to students.  We have attempted to address the problem of student learning of quantum mechanics by developing PhET simulations in this subject using our research-based design principles~\cite{Adams2008a,Adams2008b}, and conducting research on their effectiveness in various contexts.

In this paper we present an overview of the PhET quantum simulations (Section II), how they are developed (Section III), examples of classroom use and studies of their effectiveness (Section IV), and insights into student thinking we have gained from conducting student interviews on these simulations (Section V).

\section{Quantum Mechanics simulations}
We have two main goals for PhET simulations: increased
student engagement and improved learning.
Simulations are specifically designed to support students in
constructing a robust conceptual understanding of
the physics through exploration. Their design is grounded in research. We draw from existing
research literature on how students learn, conceptual
difficulties in physics, and educational technology design.
We also make extensive use of student interviews
and classroom testing to explore usability, interpretation,
and learning issues, and to develop general simulation
design principles.

PhET is best known for our simulations on topics in introductory physics, such as \emph{Circuit Construction Kit}, \emph{Masses and Springs}, and \emph{The Moving Man}.~\cite{Perkins2006a,Wieman2006a,Wieman2008a}  However, the features that make these simulations effective for learning introductory physics are even more important for learning quantum mechanics.

We now have a suite of 18 simulations on various aspects of quantum phenomena.  These fall into three broad categories, illustrated in Table 1: fundamental principles of quantum mechanics, key ideas in historical experiments, and quantum principles underlying everyday life applications.  The PhET quantum simulations are available from http://www.phet.colorado.edu/quantum.  Each simulation has a web page that includes a brief description, learning goals, sample classroom activities, and ``Tips for Teachers,'' which discuss approximations made in the simulation, common student difficulties, and suggestions for classroom use.

\begin{table}[htbp]
\resizebox{\columnwidth}{!}{
\begin{tabular}{|l|l|l|}
  \hline
  \textbf{Fundamental Principles} & \textbf{Historical Experiments} & \textbf{Applications} \\
  \hline
  Quantum Tunneling & Photoelectric Effect & Lasers \\
  \hspace{0.2in} and Wave Packets & Davisson Germer: & Neon Lights \&\\
  Quantum Wave Interference & \hspace{0.2in} Electron Diffraction & \hspace{0.1in}Other Discharge\\
  Quantum Bound States & Stern-Gerlach Experiment & \hspace{0.2in}Lamps \\
  Double Wells and & Rutherford Scattering & Simplified MRI \\
  \hspace{0.2in} Covalent Bonds & Models of the & Conductivity \\
  Band Structure & \hspace{0.2in} Hydrogen Atom & Semiconductors \\
  Fourier: Making Waves & Blackbody Spectrum & Nuclear Physics \\
  \hline
\end{tabular}}
  \caption{PhET simulations on quantum mechanics}
\end{table}

\subsection{Visualization}
%
%\begin{figure}[htbp]
%  \includegraphics[width=\columnwidth]{MHAsim}
%  \caption{\textit{Models of the Hydrogen Atom} simulation.  A gun shoots photons at a Hydrogen atom.  An energy level diagram shows the transitions and a spectrometer records the colors of the emitted photons.  In this picture, the atom has just jumped from the ground state to an excited state in the Schrodinger model.}
%  \label{MHAsim}
%\end{figure}

\begin{figure}[htbp]
  \includegraphics[width=\columnwidth]{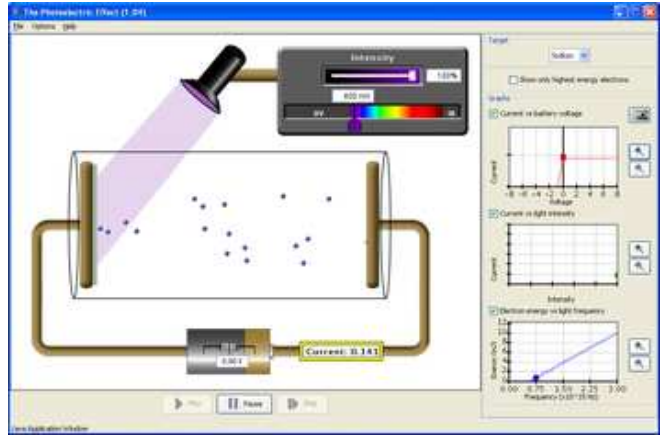}
  \caption{\textit{Photoelectric Effect} simulation.}
  %Light shines at a metal target and ejects electrons.  Students can adjust intensity, wavelength, voltage, and the type of target.}
  \label{PEsim}
\end{figure}

\begin{figure}[htbp]
  \includegraphics[width=\columnwidth]{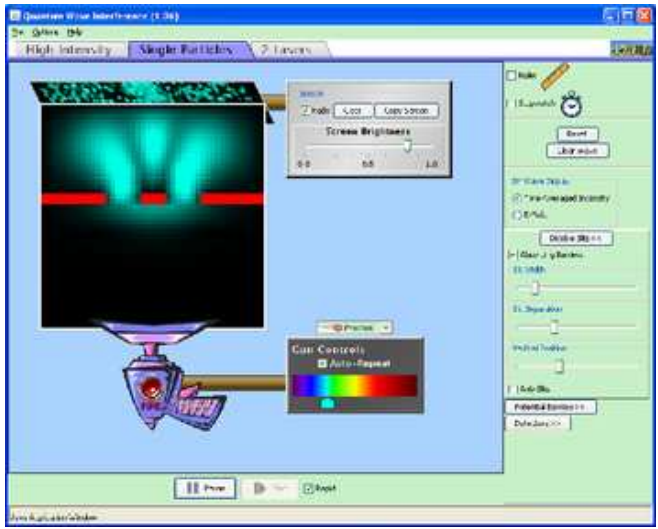}
  \caption{\textit{Quantum Wave Interference} simulation.}
  %In the ``Single Particles'' tab, a photon gun shoots individual photons at a screen, where they collapse to dots and slowly build up an interference pattern.}
  \label{QWIsim}
\end{figure}

Simulations are powerful tools for helping students visualize electrons, photons, atoms, wave interference, and other quantum phenomena that they cannot observe directly.  While students can conduct experiments on topics such as the photoelectric effect and double slit interference in many physics labs, there is much going on inside these experiments that they cannot observe.  \emph{Photoelectric Effect} (Fig.~\ref{PEsim}) allows students to watch electrons travel between the plates, helping them to build a model of why the current increases when you increase the intensity (they can see that more electrons leave the plate) but does not increase when you increase the voltage (they can see that the electrons travel faster between the plates but the number of electrons stays the same).  \emph{Quantum Wave Interference} (Fig.~\ref{QWIsim}) allows students to follow a light wave from the source and through the slits, observing it interfering with itself and collapsing into a dot on the screen.  \emph{Models of the Hydrogen Atom} (not shown) allows students to ``see'' inside atoms.

\subsection{Interactivity}
PhET simulations are highly interactive, directly coupling students' actions with the animation. Adjustment of controls results in an immediate animated response in the visual representations.  Our research with student interviews shows that this interactivity helps students engage with the content and establish cause-and-effect relationships.~\cite{Adams2008a}  Further, interactivity that allows students to switch between representations enhances students' abilities to connect multiple representations.~\cite{Adams2008a}  This interaction appears to be particularly effective for helping students construct understanding and intuition for abstract and unfamiliar quantum phenomena.  For example, in \emph{Quantum Bound States} (Fig.~\ref{twowells}), students can learn about the relationship between potential energy and wave function by clicking and dragging directly on the potential energy diagram to change the offset, height, and width of potential wells, and immediately see the effect on the shape of the wave function.
%
%\begin{figure}[htbp]
%  \includegraphics[width=\columnwidth]{QBSsim}
%  \caption{\textit{Quantum Bound States} simulation.  Students can click and drag to change the potential and select different energy levels, observing how the wave function changes.}
%  \label{QBSsim}
%\end{figure}

\subsection{Context}
\begin{figure}[htbp]
  \includegraphics[width=\columnwidth]{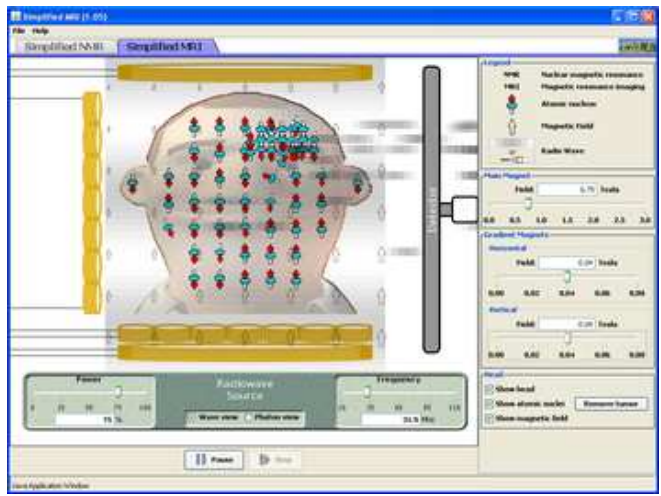}
  \caption{\textit{Simplified MRI} simulation.}
  %  Gradient magnets produce a spatially varying magnetic field, and the frequency of the radio source can be adjusted to flip the spins in different parts of the head and detect tumors.}
  \label{MRIsim}
\end{figure}

The focus on real-world contexts and applications that is a characteristic of nearly all PhET simulations is particularly helpful in grounding quantum mechanics in students' everyday experiences.  For example, \emph{Simplified MRI} (Fig.~\ref{MRIsim}) enables students to learn about nuclear spin and energy splitting in the context of MRI.  \emph{Neon Lights and other Discharge Lamps} (not shown) enables students to see how neon lights work based on the concepts of atomic energy levels, energy transfer from electrons to atoms, photon emission, and atomic spectra.  Putting basic concepts in these real-world contexts helps students appreciate the relevance of the physics, and working through how MRIs and discharge lamps work also has the potential to help them understand the physics behind these applications.

%
%\begin{figure}[htbp]
%  \includegraphics[width=\columnwidth]{NLDLsim}
%  \caption{\textit{Neon Lights and other Discharge Lamps} simulation.  A heated plate emits electrons, which are accelerated across a voltage to excite the gas of atoms inside the tube so they emit colored light.}
%  \label{NLDLsim}
%\end{figure}

\subsection{Taking advantage of the Computer}
\begin{figure}[htbp]
  \includegraphics[width=\columnwidth]{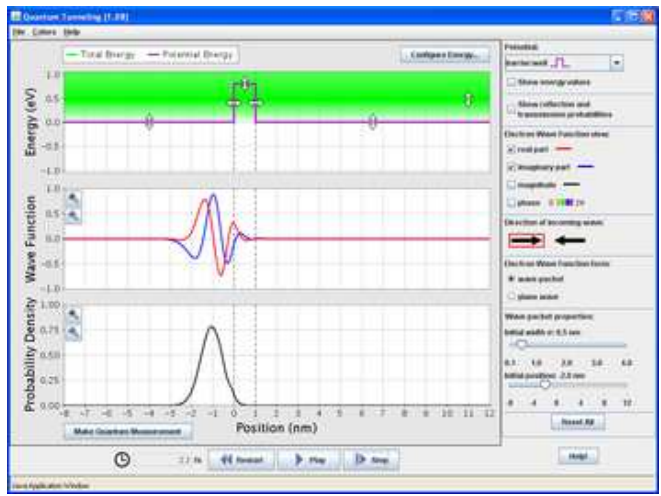}
  \caption{\textit{Quantum Tunneling and Wave Packets} simulation.}
  %  A wave packet approaches an adjustable barrier, through which it will tunnel.}
  \label{QTsim}
\end{figure}

Many of the quantum simulations take advantage of the power of computers to quickly do complex calculations without exposing the user to the details.  Thus, students can explore quantum tunneling and quantum wave interference qualitatively and focus on understanding the concepts without getting bogged down in the math.  This has the potential to radically transform the way quantum mechanics is taught because it allows the instructor to focus on the problems that are most important for students to understand rather than on the problems that are easiest to calculate.  For example, while plane waves are certainly easier to calculate than wave packets, we have found that plane waves are actually much more difficult conceptually for students to understand.~\cite{McKagan2007e} \emph{Quantum Tunneling and Wave Packets} (Fig.~\ref{QTsim}) allows us to begin our instruction on tunneling with wave packets, so that students can visualize an electron as a slightly-but-not-completely delocalized object that approaches a barrier, interacts with it, and then partially reflects and partially transmits.  This is not only much easier to visualize and understand than a wave packet spread over infinite space interacting with a barrier for all time, but also more physically accurate.

Simulations provide a unique tool for exploring time dependence in a way that is impossible in print media, helping students to see how quantum phenomena evolve and change in time.  In \emph{Models of the Hydrogen Atom}, \emph{Neon Lights and Other Discharge Lamps}, and \emph{Lasers}, students can observe atoms absorbing and emitting photons.  In \emph{Quantum Tunneling and Wave Packets} and \emph{Quantum Bound States}, students can observe how wave functions change in time, exploring, for instance, the interchange between real and imaginary parts, the oscillation of superposition states, and the collapse of the wave function when a position measurement is made.

\section{Developing Research-Based Sims}

\begin{figure}[htbp]
  \includegraphics[width=\columnwidth]{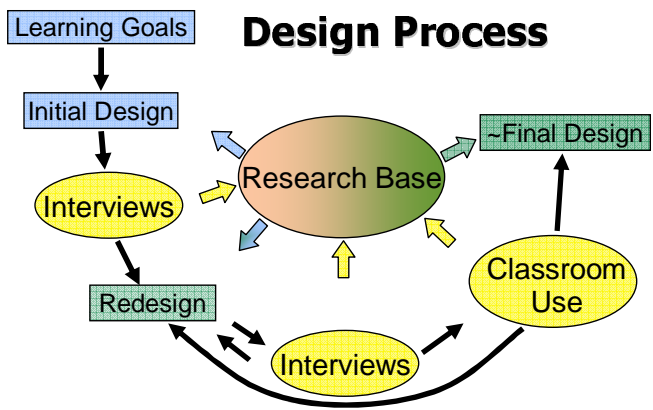}
  \caption{The PhET Design Process}
  \label{design}
\end{figure}

Fig.~\ref{design} illustrates the design process for creating PhET simulations.  The design cycle starts with content and student interface experts creating a detailed initial layout based on the learning goals of the simulation and the research base, including education and cognitive science research and the PhET design guidelines~\cite{design}.  After creating an initial version of the simulation that all team members feel is clear, accurate and engaging, we conduct student interviews to determine whether students can understand how to use the simulation and achieve the learning goals.  These interviews always reveal interface weaknesses, resolve interface questions that were not agreed upon by the team, and often reveal pedagogically undesirable (and occasionally unexpected desirable) features and subtle programming bugs. Subsequent revisions are made, and if they are extensive, a further set of interviews are conducted. These interviews are not only used to improve the particular simulation but continue to improve our research base.  After interviews establish that the desired engagement and learning is being achieved, the simulation is used in a classroom setting where student use is observed and informally evaluated.

\subsection{Building on Previous Research}
Research on how people learn~\cite{Bransford1999a} demonstrates that students learn by actively constructing their own understanding, building on their prior knowledge.  Further, experts and novices think about subjects differently.  Experts build an organized structure of knowledge that allows them to monitor and reflect on their own understanding and focus on the underlying concepts.  Novices often don't know what to focus on and get caught up in details that experts view as irrelevant.  Because working memory is limited, education should focus on essential features to reduce cognitive load.

PhET simulation design incorporates this research in many ways. Visualization and interactivity help students construct mental models. Putting physics in familiar real-world contexts helps students relate new concepts to prior knowledge. Simulations eliminate extraneous details that are unavoidable when working with real equipment, such as the color of the wires or details of how the variable voltage supply works in the photoelectric effect experiment.

We reduce cognitive load and help students construct their own understanding by starting simulations in simple states, allowing students to gradually work up to exploring more advanced features.  For example, many simulations include several tabs (e.g. Figs. \ref{QWIsim}-\ref{MRIsim}), where the first tab focuses on the basic ideas, and later tabs include more complex ideas.  In \emph{Neon Lights and Other Discharge Lamps} and \emph{Lasers}, the first tab allows students to explore the behavior of a single atom before exploring a gas of many atoms.  \emph{Fourier: Making Waves} starts up with only a single non-zero Fourier component and an invitation to add more, so that students can build up complex patterns at their own pace rather than trying to make sense of a pre-existing pattern.  These designs are based on both education research about how students learn~\cite{Bransford1999a}, and our own research showing that when we start simulations in more complex states, students become overwhelmed~\cite{Adams2008a,Adams2008b}.

Research on faculty adoption of research-based curriculum~\cite{Henderson2007a} demonstrates that instructors rarely adopt a curriculum as is, but tend to adapt it to suit their local circumstances.  At the same time, instructors need guidance on the essential features of a curriculum to help them adapt it effectively.  To enable adaptation, we design the simulations to be open-ended and general-purpose, so that each can be used in many different ways to achieve many different learning goals.  To assist instructors in using the simulations effectively, we provide guidelines for developing guided inquiry activities~\cite{guidelines} and ``Tips for Teachers'' with guidance on the use of individual simulations.  In addition, we provide a database of activities including lesson plans, lecture notes, and homework.~\cite{activities}  The database includes activities developed by the PhET team as well as those contributed by teachers.

Simulation design is also based on research into student understanding of the specific content area of the simulation. There has been some previous research on student understanding of quantum mechanics, which we have incorporated into the design of the quantum simulations.  For example, research on student learning of the photoelectric effect shows that students often have difficulty interpreting the circuit diagram, drawing qualitatively correct I-V graphs, distinguishing the effects of changing intensity and changing wavelength, and recognizing that electrons are ejected by the light rather than by the voltage.~\cite{Steinberg1996a,McKagan2007b}  We designed the \emph{Photoelectric Effect} simulation to address each of these difficulties.  The circuit is shown as a cartoon-like physical picture, rather than as an abstract diagram, and the variable voltage supply is illustrated as a battery with a slider.  Students can interactively create I-V graphs.  They can change the intensity, wavelength, and voltage, and immediately see the effects of each.   In our modern physics class, we use interactive lecture demonstrations and homework designed to use these features of the simulation to address known student difficulties.  This curriculum has been shown to be very effective at helping students understand the photoelectric effect.~\cite{McKagan2007b}

Research on student learning about quantum tunneling and wave functions shows that students often mix up wave function and energy.~\cite{Ambrose1999a,Wittmann2005a,McKagan2006a}  One possible cause of this confusion is that instructors and textbooks often draw both on the same graph.  In \emph{Quantum Tunneling and Wave Packets} and \emph{Quantum Bound States}, we show the two quantities on separate graphs.  \emph{Quantum Bound States} has also been designed to address research showing that students often have difficulty relating the shape of the wave function to the shape of the potential~\cite{Ambrose1999a} by allowing students to interactively explore the relationship between the two for a wide variety of potentials.

\subsection{Observations of Students}
In addition to addressing student difficulties seen in the literature, we often design simulations to address difficulties we see during observations of students when taking field notes in lecture and problem-solving sessions.

While there are many existing simulations on double slit interference of electrons, none address what we observed to be the biggest problem for students in understanding this phenomenon: visualizing the behavior of the electrons in between the slits and the screen.  Standard instruction often shows the pattern on the screen and assumes that students will know how to interpret this pattern, filling in the gap of the wave interference that must have created it.  Our experience interacting with students in a variety of contexts indicates that students need help constructing a model of how the electrons create this pattern.  Thus, in \emph{Quantum Wave Interference}, we show an electron as a particle-like wave packet approaching the slits and interfering with itself before collapsing to a dot on the screen.

When teaching the Davisson Germer experiment, we observed in problem-solving sessions that many students did not understand the main point of the experiment after instruction.  When instructors asked them to explain the purpose of this experiment, students remembered that electrons were only detected at certain angles, but could not explain why.  They viewed the electrons as particles that happened to bounce off at certain angles for some reason they could not understand, rather than recognizing how the observations could be explained by the wave nature of electrons.  Although \emph{Quantum Wave Interference} was not designed to address this difficulty, we found that it could be used to do so.  In working with students one-on-one, we found that if we set up an array of barriers to represent atoms and demonstrated how electron wave packets aimed at this array reflected and interfered such that there were intensity maxima at certain angles and minima at other angles, students immediately responded with expressions like, ``Oh, it's interfering like a wave!'' and were then able to correctly explain the purpose of the experiment.  While this method was effective in helping students develop a correct explanation, the array of barriers was tedious to construct and difficult to change.  Students often attempted to explore how changing the spacing and size of the barriers would change the pattern, but gave up quickly when they realized how hard it was to modify each barrier.  To facilitate such exploration, we developed a new simulation, \emph{Davisson Germer: Electron Diffraction}, in which an array of atoms is set up automatically and the spacing and size can be changed by moving a slider.  A classroom study demonstrating the effectiveness of this simulation will be presented in Section IVB.

\subsection{Student Interviews}
After developing an initial version of a PhET simulation, we test it in interviews in which students are directed to ``think out loud'' as they explore a simulation, either with no directions or with a simple guiding question.  These interviews help us refine the user interface and pedagogical effectiveness of simulations.  As we will discuss in Section \ref{interviews}, interviews also provide new insights into student thinking and simulation effectiveness.  In this section, we illustrate some examples of how we have used interviews to refine simulations.  Unless otherwise noted, in all examples discussed in this article, students were engaged in undirected exploration of a simulation.

Sometimes we come up with ideas that just don't work.  In the initial version of \emph{Photoelectric Effect}, we attempted to reduce students' cognitive load by starting with a ``simple'' model in which all electrons were ejected with the same energy.  We thought that as students became more comfortable with this simple model, we would then introduce the ``realistic'' model, in which electrons were ejected with a range of energies.  The simulation allowed students to switch between models with radio buttons labeled ``simple'' and ``realistic.''  In interviews every student got caught up trying to figure out the difference between these two modes, and either gave up or developed an incorrect explanation.  Further, when we used the simulation in class, many of the student questions during lectures and problem-solving sessions revolved around trying to understand the difference between the simple and realistic models. In response, we modified the simulation by replacing the ``simple'' and ``realistic'' radio buttons with a checkbox labeled ``show only highest energy electrons.''  It is unchecked by default, so that the simulation starts in the ``realistic'' model where electrons are ejected with a range of energies.  When we used the new version in class the following semester, student questions focused much more on the physics behind the simulation.

Interviews often reveal that seemingly small details can make a big difference in student understanding.  For example, $\Delta k$ and $\sigma_k$ are two common labels for the width of a wave packet in Fourier space.  Since most physicists are equally comfortable with both, the choice between the two seems irrelevant.  However, in interviews on an early version of \emph{Fourier: Making Waves} in which we used the label $\Delta k$, students referred to it as ``the change in k.''  Because this incorrect interpretation of the label seemed so plausible, they were never able to determine what it actually meant.  In interviews after we changed the label to $\sigma_k$, students initially had no idea what the label meant. They were thus more willing to to explore it and were able to determine the correct meaning.

Observations of what students attempt to do with simulations often helps determine what additional features are needed.  These features are often not obvious and could not have been foreseen by the developers.  For example, \emph{Neon Lights and other Discharge Lamps} includes an option for configurable atoms, in which students can modify the energy levels by clicking and dragging on them.  In the initial version, students tried to drag the pictures of atoms labeling the energy levels rather than the levels themselves, and got frustrated when this didn't work.  In response, we modified the simulation to allow students to click and drag the atoms as well as the levels.

Unfortunately, interviews sometimes reveal problems that we do not know how to solve.  We have found that students recognize when the scale is unrealistic and don't attempt to attribute meaning to the relative size of objects such as for the macroscopic images of electrons, photons, and atoms in \emph{Photoelectric Effect}, \emph{Neon Lights and other Discharge Lamps}, and \emph{Lasers}.~\cite{Adams2008a}  However, we have not found any method that communicates clearly to students when the scale changes within a simulation.  For example, in \emph{Quantum Wave Interference}, students can choose to view interference of photons, electrons, neutrons, or Helium atoms, all of which exhibit wave properties at vastly different time and distance scales.  We initially indicated the change in scale only by changing the units on the ruler and stopwatch, but students either did not notice or did not know how to interpret this change.  We then tried adding a feature in which a clock with a note that says ``slowing down time'' and/or a magnifying glass with a note that says ``zooming out'' appears when the time/distance scale changes, but students did not know how to interpret this either.  We left this feature in as a reminder, because while it didn't help students, it also didn't hurt.  Students can still learn many other things from the simulation without recognizing the change in scale, and if instructors want students to notice this change, they can point it out explicitly or incorporate this idea into homework activities.

The issue of changing scale also caused problems in an early version of \emph{Models of the Hydrogen Atom}, which allowed students to explore both atomic spectra and Rutherford scattering by shooting light and alpha particles at an atom.  However, because light interacts with the electrons and alpha particles interact with the nucleus, these two processes occur at very different scales.  Initially we tried to gloss over this fact by showing both at an intermediate scale.  This led to a great deal of confusion in interviews.  Students thought the alpha particles were interacting with the electrons and often drew incorrect conclusions, such as that alpha particles are negatively charged.  After these interviews we decided that we were trying to show too many different things in a single simulation, and split the alpha particle feature into a separate simulation called \emph{Rutherford Scattering}.

\section{Classroom Use and Testing}

\subsection{Examples of Classroom Use}
Most of the quantum simulations were developed for use in a reformed large-lecture modern physics course for engineering majors.~\cite{McKagan2007a}  This course used the simulations in a variety of contexts including: general visual aids; interactive lecture demonstrations, where we demonstrated key phenomena and asked students to make predictions about the behavior of simulations using clickers; and homework, which guided students through exploration of simulations.  Our course material is available from the PhET activities database~\cite{activities} (search for author ``McKagan'') or from our modern physics course archive~\cite{2130}.

An example of a visual aid is our use of \emph{Quantum Wave Interference} in lecture to demonstrate how the double slit experiment shows that light must be both a wave that goes through both slits and a particle that hits the screen at a single location.~\footnote{http://phet.colorado.edu/new/teacher\_ideas/view-contribution.php?contribution\_id=198}  This lecture led to an unexpected onslaught of deep, fundamental questions that took up nearly an entire class period.  Student questions included:
\begin{itemize}
  \setlength{\itemsep}{1pt}
  \setlength{\parskip}{0pt}
  \setlength{\parsep}{0pt}
    \item How can it be such a huge blob and then be detected in one place?
    \item Is it that we just don't know where it is or is it really spread out in space?
    \item What does it take for a photon to collapse to a single point?  How does that happen?
    \item In real life can you really turn it down so low that you only have one photon coming out at a time?
    \item How big is a photon?  Can it be a meter wide?
\end{itemize}
These questions are similar to those asked by the founders of quantum mechanics as they worked out the meaning of this new theory.  Student difficulties often do not reflect the historical questions of scientists because students struggle with much more basic questions.  In this case, we argue that the visualization provided by the simulation allowed students to see the heart of the issue and ask deep questions earlier in the learning process.

\begin{figure}[htbp]
  \fbox{\includegraphics[width=\columnwidth]{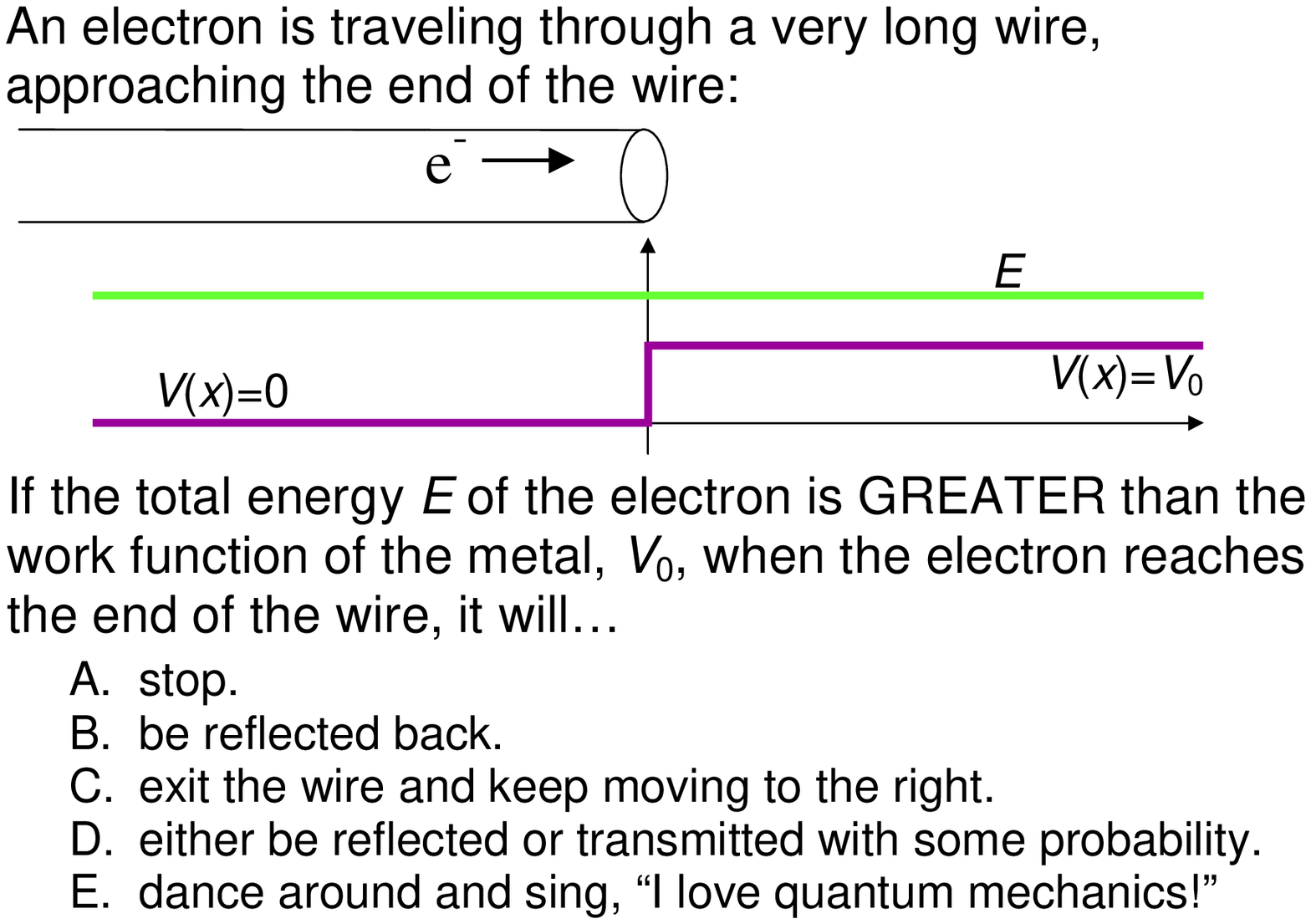}}
  \caption{\label{reflection}Sample interactive lecture demo question with \emph{Quantum Tunneling and Wave Packets}.  The correct answer is D.}
\end{figure}

Fig.~\ref{reflection} illustrates the use of a simulation for an interactive lecture demonstration.  This is a typical question in which students discuss with their neighbors and then select an answer using clickers.  After asking a few students to give explanations for their answers to the whole class, we use \emph{Quantum Tunneling and Wave Packets} (Fig.~\ref{QTsim}) to demonstrate what actually happens.  Students always point out that the simulation in fact shows the wave being partially reflected and partially transmitted, rather than reflected OR transmitted, as in the correct answer D.  After a class discussion of this discrepancy, we use the ``Make Quantum Measurement'' button in the simulation to demonstrate that after it is measured, the electron is always reflected or transmitted, never both.

Another way to use simulations is in homework.  In our modern physics class, students work through a series of questions using \emph{Lasers} to build up an understanding of how a laser works.~\footnote{http://phet.colorado.edu/new/teacher\_ideas/view-contribution.php?contribution\_id=56}  The homework starts with basic questions about absorption and spontaneous and stimulated emission, works through the steps of building a laser and troubleshooting a broken laser, and ends with essays on why a population inversion is necessary to build a laser and why this requires atoms with three energy levels instead of two.  Most students are able to give correct and thorough explanations in these essays.

\subsection{Classroom testing of simulation effectiveness}
We have conducted several studies in our reformed modern physics course to test the effectiveness of simulations and other aspects of the course.

The most extensive testing of classroom use of a specific quantum simulation has involved \emph{Photoelectric Effect}.  In a recent study we showed that with our curriculum that included both interactive lectures and homework using the simulation, learning much greater than with either traditional or previous reformed instruction.~\cite{McKagan2007b}  For example, on an exam question about whether increasing the voltage between the plates would lead to electrons being ejected when the light frequency was too low, an average of $83\%$ of students answered correctly with correct reasoning in the courses using the simulation, compared to $20\%$ of students in a traditional course and $40\%$ of students in a traditional course supplemented by a research-based computer tutorial.

In the course as a whole, where simulations were used extensively in all the ways discussed in Section IVA, we found high learning gains (measured by the Quantum Mechanics Conceptual Survey~\cite{QMCS}) and a lack of shift in beliefs about physics (measured by the Colorado Learning Attitudes about Science Survey~\cite{Adams2006a}).  In contrast, in the course for engineering majors the semester before our reforms and in the corresponding course for physics majors, there were low learning gains and large negative shift in beliefs.~\cite{McKagan2007a}  While we made many reforms in this course, the simulations played a large role in all of them, and likely contributed to the improved learning and beliefs.

%\begin{figure}[htbp]
%  \includegraphics[width=\columnwidth]{useful}
%  \caption{\label{useful}Student ranking of usefulness of simulations for learning. ($N = 173$)}
%\end{figure}
%
Student perceptions provide a further indicator of simulation effectiveness.  On the end-of-term survey for our modern physics course ($N = 173$), the average student ranking of the usefulness of the simulations for their learning on a scale of 1 (not useful) to 5 (a great deal) was 4.0, close to the highest ranked aspect of the course.  The usefulness rankings for other aspects of the course ranged from 3.2 (the textbook) to 4.3 (the posted lecture notes).  Students also had the opportunity to make comments about the simulations in the survey, and $35\%$ ($N = 61$) chose to do so.  Of these comments, $80\%$ were positive comments about the usefulness of the simulations, for example:
\begin{itemize}
  \setlength{\itemsep}{1pt}
  \setlength{\parskip}{0pt}
  \setlength{\parsep}{0pt}
    \item Great sims, I can't imagine QM without them.
    \item The simulations were crucial in the learning process.
    \item The simulations were the best part of class, they practically answer physics questions all by themselves. I would recommend continuing to develop these and add more. Without these I think I would have been lost in the course.
    \item I definitely not only enjoyed the simulations, but I'd go as far to say that the simulations taught me the most about the course because I could really visualize the inner workings of the physics processes that were going on.
    \item I thought the simulations were great. It helped me to gain intuition about the topic. This is especially useful in quantum mechanics where it is not normally possible to directly observe the described phenomena.
\end{itemize}
Other types of comments about the simulations included pointing out: that the simulations need guidance to be useful ($13\%$); that the simulations were incorrect or not useful ($8\%$); specific technical problems ($7\%$) (most of these have since been resolved); and that the simulations are not experiments ($3\%$).\footnote{The percentages add up to more than $100\%$ because some students made more than one type of comment.}

We also conducted qualitative observations of students in this course by taking field notes in lecture and problem-solving sessions and conducting regular interviews with six students.  In all these observations, we consistently saw that for topics where we used simulations, students developed extremely vivid mental models.  For example, when we asked students in problem-solving sessions and interviews about topics related to simulations, they gave animated responses easily and without much time for thought.  On exam questions on topics such as the photoelectric effect, discharge lamps, and lasers, students gave vivid, detailed responses, often referring to the simulations explicitly and correctly remembering minute details.  On the other hand, when we asked students about other topics not related to simulations, such as models of the atom or infinite square wells (before we developed simulations on these topics), students had to think for a long time, attempting to retrieve memorized facts, and often mixed up important details.

In another classroom study, we set out to determine the prevalence of student difficulties with the Davisson Germer experiment discussed in Section IIIB, and to evaluate the effectiveness of the simulation in addressing these difficulties.  This study was conducted in a modern physics course for physics majors that used much of the same curriculum as the course for engineering majors discussed elsewhere in the paper, but had a very different population of students.  In general, the physics majors seemed to pick things up more quickly and had fewer of the difficulties observed among the engineering majors.

\begin{figure}[htbp]
  \includegraphics[width=\columnwidth]{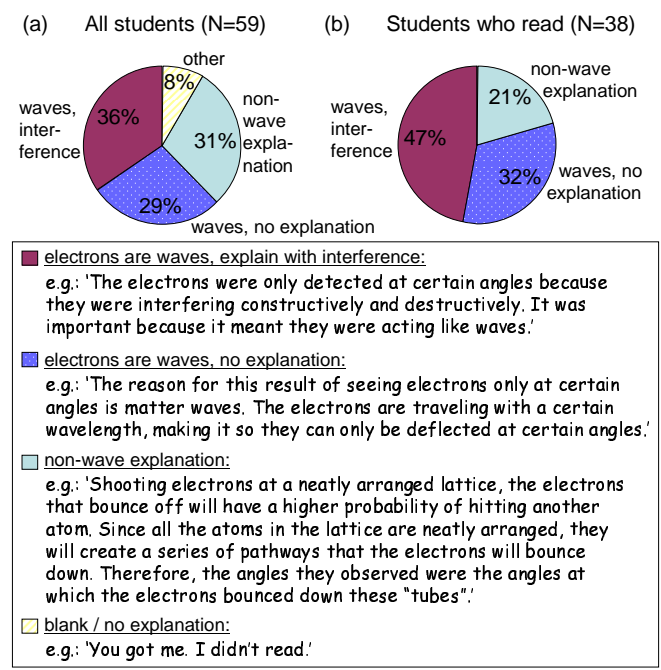}
  \caption{\label{DGresults}Student responses to reading quiz on Davisson Germer experiment.}
\end{figure}

In the class before students were expected to read the relevant section of the textbook (Knight 24.4~\cite{Knight2004a}), they were told that there would be a reading quiz on the Davisson Germer experiment in the next class.  This reading quiz was emphasized much more than usual, to ensure that students would do the reading.  In the quiz, students were asked to answer to the following question: ``In the Davisson Germer experiment, Davisson and Germer shot a beam of electrons at a lattice of Nickel atoms and found that the electrons were only detected at certain angles.  Explain the reason for this result and why it was important.''  The correct answer, that the observed pattern was characteristic of an interference pattern and therefore it showed that electrons behave as waves, is illustrated by the first sample student response in Fig.~\ref{DGresults}.  As shown in Fig.~\ref{DGresults}a, only about a third of students gave this type of answer.  Another third said that electrons are waves but did not explain this result in terms of interference, as in the second sample response.  Most of the remaining third gave an incorrect explanation that did not involve wave behavior at all, as in the third sample response.

Since it is possible that these poor results were the result of students not doing the reading, a clicker question in the next class asked students whether they did the reading on the Davisson Germer experiment.  Students were promised that their instructor would not look at their individual responses, and $35\%$ admitted to not doing the reading or not remembering whether they did or not.  Fig.~\ref{DGresults}b shows the responses to the reading quiz for the students who said they did the reading.  (Students who may have done the reading but were not present for the clicker question are not included in Fig.~\ref{DGresults}b.)  When counting only students who did the reading, the percentage of students who answered the reading quiz question correctly goes up to nearly half.  However, the remaining half either did not explain how the experiment leads to a wave model, suggesting that they had simply memorized the answer without understanding it, or gave explanations of the experiment that did not involve waves at all.

The reading quiz was followed by an interactive lecture on the Davisson Germer experiment, in which the \textit{Davisson Germer: Electron Diffraction} simulation was used in one of the two sections, and a homework in which students in both sections were asked to use the simulation to explain the Davisson Germer experiment and its application to understanding the structure of crystals.~\footnote{http://phet.colorado.edu/new/teacher\_ideas/view-contribution.php?contribution\_id=319}  On a midterm exam, students were again asked to explain the inferences that could be drawn from this experiment, but were not told what was seen in the experiment. On the exam question, $92\%$ of students correctly explained that there was an interference pattern that illustrated the wave nature of electrons.  These exam results indicate that the simulation, along with the accompanying lecture and homework, was extremely effective in helping students understand the Davisson Germer experiment.

\section{Learning from Student Interviews}\label{interviews}
The primary purpose of the think-aloud interviews that we conduct as part of our simulation design process is to find problems with the simulations in order to improve them.  However, interviews are also valuable for demonstrating the effectiveness of simulations and giving general insights into student thinking.  In this section we present some examples of what we have learned from interviews on quantum simulations.

Interviews help us determine what students can and cannot learn from each simulation.  (Details of these insights are provided in the ``Tips for Teachers'' available from the web page for each simulation.)  We have found that students can usually learn some important concepts from undirected exploration of simulations, but they can learn much more from using the simulations in conjunction with activities that guide their exploration. With undirected exploration, students can often give correct explanations of many of the concepts that the simulation is designed to teach, but do not necessarily recognize that they have learned, often because they do not understand the significance or application of the content.

For example, students with no previous instruction on Fourier analysis who explored \emph{Fourier: Making Waves} in interviews were able to give correct descriptions of Fourier analysis and explain everything in the first tab, but claimed that they did not understand the point of the simulation.  Students with no background in modern physics who explored \emph{Photoelectric Effect} were able to correctly explain how the experiment worked and the results, but did not make connections to the greater implications for the nature of light.

The simulations can be greatly enhanced by a good activity (e.g. homework, lab, or interactive lecture) that guides students' exploration towards the learning goals of the instructor and helps place the concepts within a larger context.  Many activities are available for the quantum simulations on the PhET website~\cite{activities}, some developed by PhET team members as part of our modern physics course transformation, and some contributed by users.  While we have not done studies that directly compare undirected and guided exploration of the simulations, we have measured learning gains from using guided activities in the studies discussed in the previous section, and we have seen improved student learning as a result of guidance in interviews.

\begin{figure}[htbp]
  \includegraphics[width=\columnwidth]{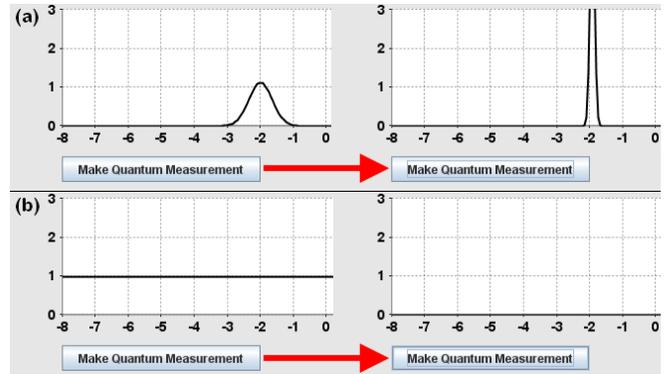}
  \caption{\label{measurement}The ``Make Quantum Measurement'' button in \emph{Quantum Tunneling and Wave Packets} measures the position of the electron in the probability density vs. position graph in (a) ``wave packet'' mode and (b) ``plane wave'' mode.}
\end{figure}

For example, in interviews on \emph{Quantum Tunneling and Wave Packets}, most students were able to explain the behavior of the simulation after they hit the `Make Quantum Measurement'' button in ``wave packet'' mode, but needed help from the interviewer to explain the behavior in ``plane wave'' mode (see Fig.~\ref{measurement}).  In ``wave packet'' mode, this button causes the probability density to collapse to a narrow packet whose position is determined randomly according to the probability density immediately before measurement.  In ``plane wave'' mode, this button causes the probability density to go to zero everywhere.  Out of six students interviewed on this simulation, three were able to explain the behavior in ``wave packet'' mode without help, two were able to explain it after a hint from the interviewer, and one never made sense out of it.  Three students who had successfully explained the behavior in ``wave packet'' mode then tried the button in ``plane wave'' mode.  All three expressed confusion over what they saw and none were able to explain it on their own.  The interviewer then asked, ``How far does a plane wave extend in space?''  All three students quickly answered that it extends over infinity, and therefore the probability of measuring it in the region shown on the screen is zero.  The results of these interviews suggest that a guided activity including the question asked by the interviewer could help students learn the effects of measurement on plane waves much more effectively than unguided exploration.

\begin{figure}[htbp]
  \includegraphics[width=\columnwidth]{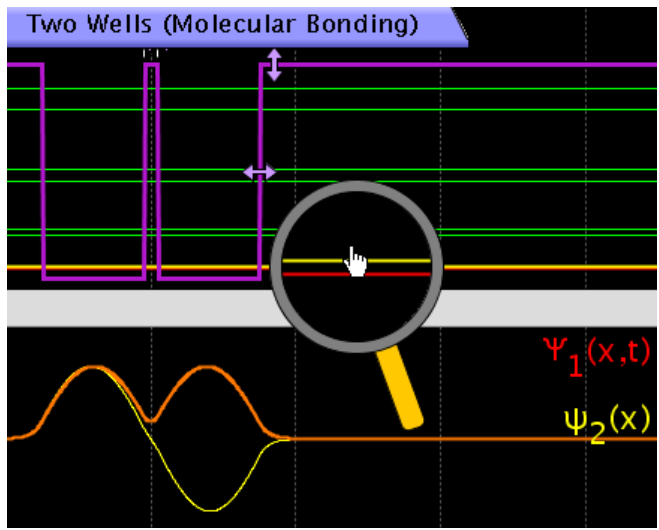}
  \caption{\label{twowells}The ``Two Wells'' tab of \emph{Quantum Bound States} (also \emph{Double Wells and Covalent Bonds}), showing the symmetric ($\psi_1$) and anti-symmetric ($\psi_2$) states.  The ``Many Wells'' tab (also \emph{Band Structure}), not shown, allows users to create an array of up to $10$ wells.}
\end{figure}

An interview on \emph{Quantum Bound States} provides an example of a student learning an advanced topic from undirected exploration of a simulation.  This simulation contains two advanced tabs that allow students to explore double and multiple wells (Fig.~\ref{twowells}).  (These tabs are also available separately as the simulations \emph{Double Wells and Covalent Bonds} and \emph{Band Structure}.)  In most of our interviews with \emph{Quantum Bound States}, the students spent so long playing with single wells that they never got to the advanced tabs, so we have only conducted one interview in which the student spent more than a few minutes playing with two wells.  This student, who had previous instruction on single wells but not double wells, was able to explain, based on his exploration of the simulation, the reason for the pairs of symmetric and anti-symmetric states for double wells: ``...because we have two wells here, so... I want to think that one is more centered around this one and the other is more around this one, and I guess we don't know which one is which, which is why they're both symmetrical around these.''  He was troubled, however, that he was unable to determine the physical interpretation of the difference between these two states.  While it is possible that this interview result was idiosyncratic, it is a valuable existence proof that it is possible for a student to learn a very advanced concept from undirected exploration of the simulation.  If one student can learn so much with so little guidance, it is likely that many students can learn this concept with a guided activity.

Interviews also help us determine the range of levels of students for which simulations are appropriate.  There are a few simulations, such as \emph{Quantum Tunneling and Wave Packets}, \emph{Quantum Bound States}, \emph{Double Wells and Covalent Bonds}, \emph{Band Structure}, and \emph{Davisson Germer: Electron Diffraction}, that require a basic knowledge of the phenomena being illustrated and therefore do not appear to be effective for students who have not had any instruction on the relevant topics.  Other simulations that one might imagine are too advanced have proven to be surprisingly effective for a wide range of students.  For example, several simulations have been used successfully in lecture demos and homework in courses for non-science majors, including \emph{Nuclear Physics}, \emph{Conductivity}, and \emph{Semiconductors} in ``The Physics of Everyday Life,'' and \emph{Fourier: Making Waves} in ``Sound and Music.''  In interviews, after a half hour of unguided exploration, students with no science background have been able to give good qualitative explanations of the physics behind simulations such as \emph{Quantum Wave Interference}, \emph{Lasers}, \emph{Neon Lights and Other Discharge Lamps}, \emph{Photoelectric Effect}, \emph{Nuclear Physics}, and \emph{Semiconductors}.  Finally, a PhET team member's 9-year-old son enjoys playing with \emph{Lasers} and \emph{Models of the Hydrogen Atom}, and has figured out much of the basic physics behind these simulations.  For example, he can explain how the photons change the energy levels and how to change the separation of energy levels to match energy of light to get lasing in \emph{Lasers}.

In interviews with students who have had previous instruction on the topics covered by the simulations, we find that the visual representations can help students address incorrect models that would otherwise be difficult for an instructor to detect.  For example, one student, upon seeing the wave packet representing a photon in \emph{Quantum Wave Interference}, said, ``Until now, I thought that, if I were to represent one particle, it would just be one thin line going up.  I did not know that it would be like, all over here.''  When the interviewer asked why he thought it would be one thin line, he described his instructor drawing a series of thin lines [wave fronts] and referring to a line as ``this one wave.''  Another student initially predicted that if you moved the slits further from the screen the separation between the interference fringes would decrease, and was able to use the simulation to correct his prediction and develop an explanation for why the separation actually increases.

Interviews on \emph{Models of the Hydrogen Atom} provide a further example of a simulation uncovering an incorrect model developed from previous instruction.  In these interviews, some students described the Plum Pudding model as a cloud of negative charge filled with little specks of positive charge, rather than the other way around.  The use of the word ``cloud'' suggests that these students are mixing up the Plum Pudding model with the Schrodinger model, in which the electrons are often described as a cloud of negative charge.  These students initially thought that the electron in the simulation was a proton, but were eventually able to identify it correctly by using the legend or by comparing it to the electrons in other models.

\begin{figure}[htbp]
  \includegraphics[width=\columnwidth]{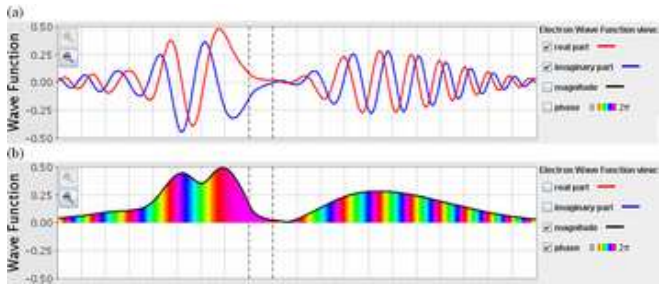}
  \caption{\label{wavefunction}Representations of the wave function in \emph{Quantum Tunneling and Wave Packets}: (a) real and imaginary parts and (b) magnitude and phase.  In interviews students can make sense of (a) but struggle with (b).}
\end{figure}

The development of \emph{Quantum Wave Interference}, \emph{Quantum Tunneling and Wave Packets}, and \emph{Quantum Bound States} illustrates what we have learned about what representations of wave functions are most conducive to student learning. Our team put a lot of thought into how to represent quantum wave functions (Fig.~\ref{wavefunction}) in these simulations.  Most textbooks only show plots of the real part of the wave function, but also discuss the imaginary part of the wave function.  The magnitude and phase may be discussed in a junior level quantum mechanics course, but usually not in a sophomore level modern physics course.  Most non-PhET simulations of wave functions use a ``phase color'' representation in which a curve representing the magnitude of the wave function is filled in with colors representing the phase (Fig.~\ref{wavefunction}b).

In observations and interviews in many contexts, we noticed that students often asked about the meaning of the imaginary part of the wave function, but never about the real part.  Further, students often forgot about the imaginary part entirely, or said that you only need to consider the real part when squaring the wave function.  We hypothesize that students overestimate the importance of the real part and underestimate the importance of the imaginary part for two reasons.  First, the unfortunate choice of words ``real'' and ``imaginary'' naturally leads to the idea that one is more ``real'' than the other in the common English sense of the word, when in fact both components are on equal footing mathematically.  Second, the fact that many textbooks illustrate only the real part (but label it as $\psi$) may encourage students to focus only on this part of the wave function.

To address this problem, we illustrate both the real and imaginary parts on equal footing in the simulations (Fig.~\ref{wavefunction}a).  We suspected that the real and imaginary parts of the wave function would be easier for students to understand than the magnitude and phase, because these representations relate more easily to what students typically calculate and to familiar sine and cosine waves.  However, we also included options to show the magnitude and phase color for completeness.

In interviews on \emph{Quantum Wave Interference}, one student commented that he did not understand real and imaginary numbers, and one student who wondered why the imaginary part didn't look different from the real part until he paused the simulation and could see that they were out of phase.  Aside these two students, whose confusion stemmed more from their expectations than from the simulation, students did not express any confusion over the real and imaginary representations of the wave function in interviews on \emph{Quantum Tunneling and Wave Packets} and \emph{Quantum Wave Interference}.  Several students also learned important concepts by playing with the real and imaginary views.  For example, students learned from the simulation that the real and imaginary parts were 90 degrees out of phase, and that the real and imaginary parts add up to a constant probability density in an energy eigenstate even though each individual component changes in time.

On the other hand, the ``phase color'' representation caused significant problems for most students.  In interviews on \emph{Quantum Wave Interference}, three out of five students interviewed explored this view.  None of the three made any comments on it on their own, aside from one student who said it hurt his eyes, so the interviewer asked them what it was showing.  One student said it was ``some sort of frequency type of thing'' and speculated that teal would constructively interfere with teal and destructively interfere with the opposite of teal.  Another stared at the screen in confusion for a minute, and then described it as ``some sort of representation of both the real part and the imaginary part'' showing that ``pink is areas of high real part and low imaginary part or something?''  Another student was unable to give any explanation.  When the same three students were interviewed later on \emph{Quantum Tunneling and Wave Packets}, the two who had given explanations in earlier interviews did not comment on phase view again.  The student who had been unable to give any explanation remembered that this view had been used in his quantum course, but still could not explain what it meant.  Of three additional students who were interviewed on \emph{Quantum Tunneling and Wave Packets} but not \emph{Quantum Wave Interference}, two expressed frustration over the phase view and were unable to explain it, and the third, when asked to explain it, said only that it showed ``something about wavelength.''  When given a choice, none of the students spent much time in phase mode, returning quickly to real or magnitude mode after answering the interviewer's questions.

``Phase color'' is still an option in the simulations for instructors who would like to explicitly teach the use of this representation or use activities developed for other simulations, but we recommend caution in its use.

\section{Conclusion}

In summary, PhET quantum simulations are designed to address previously-known student difficulties in quantum mechanics, as well as many new student difficulties uncovered as a result of our research.  The key features of PhET simulations - visualization, interactivity, context, and effective use of computations - are particularly effective for helping students understand the abstract and counterintuitive concepts of quantum mechanics.  Our research has shown these simulations to be effective in helping students learn, and has revealed new insights into how students think about quantum mechanics.

\section{Acknowledgments}
We thank Noah Finkelstein, Wendy Adams, and the rest of the PhET team and the Physics Education Research Group at the University of Colorado.  We gratefully acknowledge the NSF, the Hewlett Foundation, the Kavli Foundation, and the University of Colorado for providing the support to develop the simulations, and to make them freely available to all educators and students.

%\bibliography{../bibliographies/PER}

\bibliographystyle{apsrev}
\end{document}